\documentclass[]{pasj02} 
\usepackage[switch,mathlines]{lineno} 

\begin{document} 

\title{Detection of Doppler velocity differences between ions and neutrals in an erupting prominence}

\author{
 Yuwei \textsc{Huang},\altaffilmark{1}\altemailmark
 \email{lelouchkako@gmail.com} 
 Kiyoshi \textsc{Ichimoto},\altaffilmark{2,1}\altemailmark $^{,\dag}$
 \email{ichimoto@kwasan.kyoto-u.ac.jp}
 and
 Tetsu \textsc{Anan},\altaffilmark{3}\altemailmark $^{,\dag}$ 
 \email{tanan@nso.edu}
}

\altaffiltext{1}{Astronomical Observatory, Kyoto University, Kamitakara, Gifu 506-1314, Japan}
\altaffiltext{2}{College of Science and Engineering, Ritsumeikan University, Kusatsu, Shiga 525-8577, Japan}
\altaffiltext{3}{National Solar Observatory, 22 Ohi`a Ku, Makawao, HI 96768, USA}

\footnotetext[$\dag$]{ORCID: 0000-0001-8689-3564, 0000-0001-6824-1108}



\maketitle

\begin{abstract}
We performed a spectroscopic observation of an erupting prominence occurred
on the solar limb on 2015 May 8th in He$\,\emissiontype{I}$ 7065 \rm{\AA}, 
O$\,\emissiontype{I}$ 7772 \rm{\AA} triplet and Ca$\,\emissiontype{II}$ 8498 \rm{\AA} lines  
to investigate differences in the Doppler velocity between ions and neutrals 
in a plasma strongly accelerated by the Lorentz force. 
We found that the ion-neutral velocity difference between Ca$\,\emissiontype{II}$ and He$\,\emissiontype{I}$ 
reached an order of 15 $\rm{km\, s^{-1}}$. 
On the other hand, the velocity difference between 
Ca$\,\emissiontype{II}$ and O$\,\emissiontype{I}$ was significantly 
smaller than that between Ca$\,\emissiontype{II}$ and He$\,\emissiontype{I}$.
This result can be interpreted as the formation of O$\,\emissiontype{I}$ 7772  \rm{\AA} lines
in the erupting prominence is mainly contributed by the recombination 
from O$\,\emissiontype{II}$ ions through charge transfer with hydrogen atoms, 
resulting in a behavior close to ions.
According to an order estimate of the collisional friction 
among He$\,\emissiontype{I}$ atoms and protons,
the observed velocity difference between Ca$\,\emissiontype{II}$ and He$\,\emissiontype{I}$ implies 
the acceleration of the eruption reaching about 150 times of solar gravity.
We propose a new method to evaluate the ionization degree of hydrogen
from the velocity differences observed 
in Ca$\,\emissiontype{II}$, He$\,\emissiontype{I}$ and O$\,\emissiontype{I}$ lines.
\end{abstract}


\section{Introduction}

In many astrophysical systems, plasma is partially ionized, where neutral atoms feel the Lorentz force indirectly through the collisional friction with charged particles. 
Therefore neutrals may diffuse across the magnetic field, and there exist velocity differences between ions and neutrals. 
This diffusion process as called ambipolar diffusion in astrophysics plays a key role in modifying important physical processes such as magnetic reconnection, instabilities, plasma heating, and transport of angular momentum through the magnetic field in the formation and evolution of circumstellar disks and stars \citep{Leake2012, Khomenko2012, Khomenko2014, Tomida2015, Hillier2016, Snow2020}. 
It will also cause a non-uniformity of element abundance in plasma structures 
due to the differential diffusion depending on the atomic mass \citep{Gilbert2007}.

Direct observational evidence of the diffusion of the neutral gas 
in astronomical bodies is 
of a crucial importance to validate this theoretical prediction. 
Previous observational studies have presented Doppler velocity differences between neutral and ions for quasi-static phenomena. 
\citet{Khomenko2016} detected 
Doppler-velocity differences between ions and neutrals of the order 
of $\pm 0.1$ $\rm{km\,s^{-1}}$ in a quiescent prominence
with spectral lines He$\,\emissiontype{I}$ 10830 \rm{\AA} and Ca$\,\emissiontype{II}$ 8542 \rm{\AA}. 
\citet{Wiehr2021} observed a quiescent prominence 
in He$\,\emissiontype{I}$ 5016 \rm{\AA} and Fe$\,\emissiontype{II}$ 5018 \rm{\AA} 
with a slit of spectrograph fixed on the prominence, 
and found the excess of ion velocity against that of neutral up to 35\%. 
\citet{Zapior2022} also observed another quiescent prominence in Ca$\,\emissiontype{II}$ H, H$\epsilon$, H$\beta$, He$\,\emissiontype{I}$ D3, H$\alpha$, and Ca$\,\emissiontype{II}$ 8542 \rm{\AA} with a slit scanning over the prominence, 
and found differences in the Doppler velocity of ions and neutrals up to $1.7 \,\rm{km\,s^{-1}}$ in optically thin part with a large velocity gradient at prominence edges. 
On the other hand, 
\newpage
\noindent \citet{Anan2017}, also targeting a quiescent prominence, 
observed spectral lines H$\epsilon$, H$\,\emissiontype{I}$ 4340 \rm{\AA}, 
Ca$\,\emissiontype{II}$ H, and Ca$\,\emissiontype{II}$ 8542 \rm{\AA}.
Although they found occasional velocity difference exceeding 3 $\sigma$ noise level, 
the difference between the same species is not significantly different 
from those between neutral atoms and ions.
They interpreted the result as motions of different parts in the prominence 
along the line-of-sight 
rather than the decoupling of neutral atoms from ionized plasma.

The velocity of neutral particles with respect to the ionized plasma was studied 
by \citet{Gilbert2002} based on a formulation of collisional friction among different species.
In a typical physical condition of quiescent prominences, 
they derived the velocity of drainage due to the gravity 
for neutral hydrogen and helium as $3.7 \,\rm{m\,s^{-1}}$ and $81 \, \rm{m\,s^{-1}}$, respectively.
Here, let's consider a simple model to recall the basic feature 
of the ion-neutral decoupling, i.e.,  
a plasma consisting of neutral particles (with mass $m_n$) 
and ionized hydrogen (proton with mass $m_p$) suspended 
against the solar gravity in the corona by a magnetic field.
Assume that the collisional friction imposed on the neutral is entirely attributed 
to the impact of protons and is proportional 
to the relative bulk velocity, $\Delta V$, between neutral and proton gas.
Provided that the friction balances with the gravity, 
then $\nu m_p \Delta V = m_n g$, 
where $\nu$ is the collisional frequency, and $g$ is acceleration of gravity.
Since $\nu$ is proportional to the number density of proton, $n_p$, and its 
thermal velocity, $V_{th}$,
we can express $\nu$ as $\nu \propto n_p V_{th} \sim n_H r \sqrt{k_BT/m_p}$ 
with $n_H$ the number densities of hydrogen, 
$r$ the ionization degree of hydrogen, 
and $k_B T$ temperature multiplied by the Boltzmann constant.
By replacing $g$ by $a$, i.e., acceleration due to the Lorentz force,
we obtain $\Delta V \propto a m_n (m_p T)^{-1/2} (n_H r)^{-1} $.
This relation implies that a larger velocity of neutral relative to ions is expected with 
a larger acceleration and heavier neutral atoms
for a given hydrogen density, temperature, and ionization degree. Therefore, observing an explosive phenomena with spectral lines 
originating from heavy neutral elements is advantageous for the detection of ion-neutral
decoupling.

In this study, we present an observation of an off-limb solar eruption 
in spectral lines He$\,\emissiontype{I}$ 7065 \rm{\AA}, O$\,\emissiontype{I}$ 7772 \rm{\AA} triplet 
and Ca$\,\emissiontype{II}$ 8498 \rm{\AA}, 
in order to investigate the decoupling of neutral atoms from plasma in a dynamic phenomenon. 
Using spectral line of heavy neutrals, i.e., neutral oxygen, 
we expected to detect a larger ion-neutral velocity difference than 
that of light neutral helium. 
Since we were observing erupting features triggered and governed by magnetic field, 
atoms suffered much larger acceleration than gravity.

In the following, details of the observation and data reduction 
are described in sections 2 and 3, 
the results on Doppler velocities in section 4,  
discussions on the results are given in section 5 and conclusion in section 6.

\section{Observation and data processing}

\subsection{Observation}

\begin{figure*}
 \begin{center}
  \includegraphics[width=16cm]{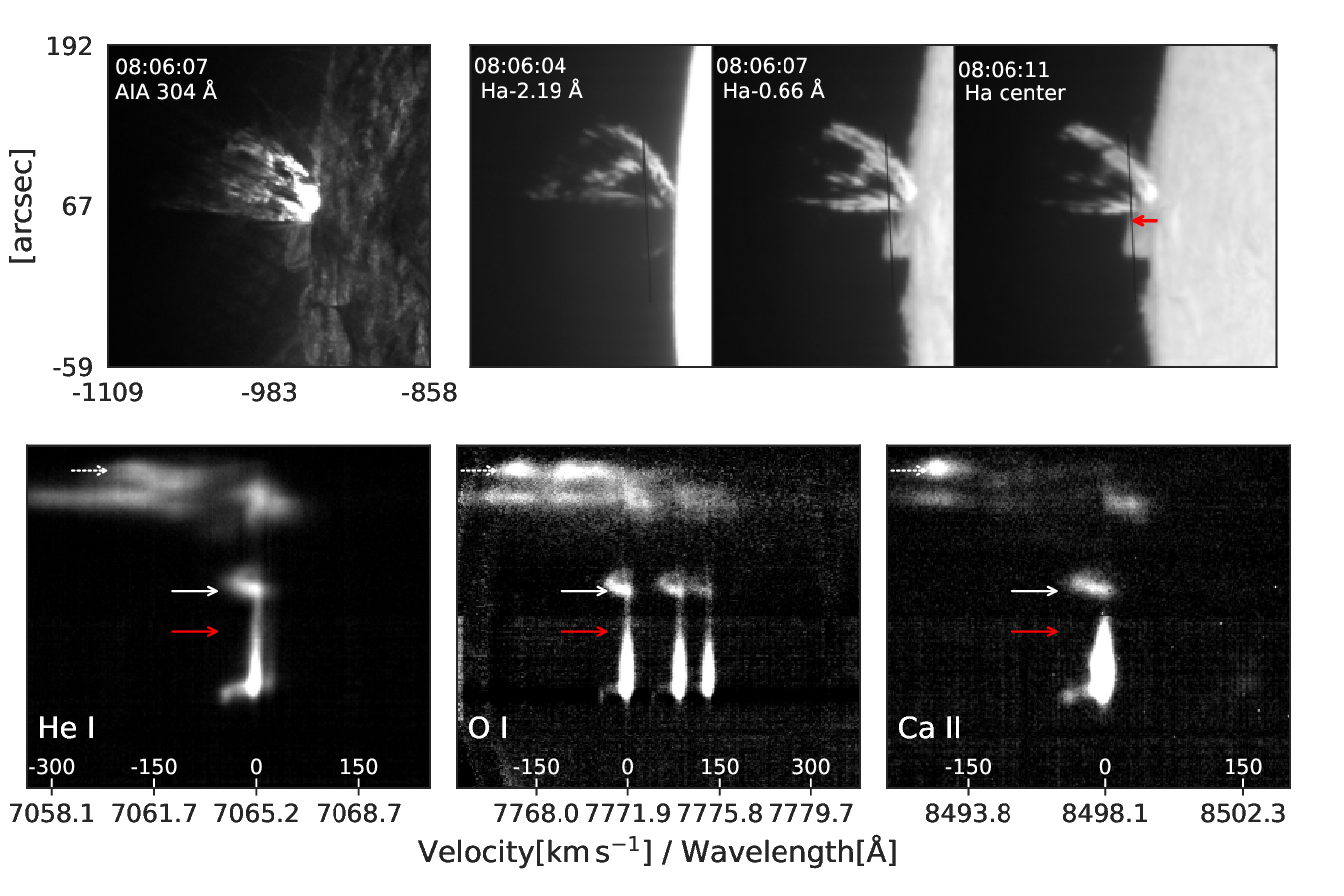} 
 \end{center}
\caption{Top left: the prominence eruption in AIA 304 \rm{\AA}; 
top right: slit-jaw images in H$\alpha-2.19$ \rm{\AA}, 
H$\alpha-0.66$ \rm{\AA} and H$\alpha$ center, 
where the oblique dark line over the prominence
shows the slit of spectrograph. 
Lower panel: spectra of He$\,\emissiontype{I}$ 7065 \rm{\AA}, 
O$\,\emissiontype{I}$ 7772  \rm{\AA} triplet and 
Ca$\,\emissiontype{II}$ 8498  \rm{\AA} from the left to the right.
These spectra were taken at 8:06 UT. 
White solid and dashed arrows mark positions 
where we took the time sequences I and II shown in figure \ref{fig:fig-seq}, respectively. 
Red arrows indicate the position of the stationary prominence in which 
we obtained the reference spectra by averaging adjacent 5 pixels.}
{Alt text: Four monochromatic images of the erupting prominence are shown in upper row.
The left image is labeled as AIA304\AA\ at 08:06:07. 
Three images in the right are labeled as H$\alpha$-2.19\AA\,  
H$\alpha$-0.66\AA\ and  H$\alpha$ center, respectively.
In the lower row, spectral image of three windows are shown 
with a label of He I, O I and Ca II from the left to the right. 
In spectral images, horizontal axis is the wavelength and vertical axis is
the spatial position along the slit.}
\label{fig:fig1}
\end{figure*}

The data in this study were acquired using the horizontal spectrograph of the Domeless Solar Telescope (hereafter DST for short, \cite{NakaiHattori1985}) 
at Hida observatory, Japan on 2015 May 8th. 
We focus on a time series of spectra where an off-limb solar eruption (N18$^{\circ}$, E90$^{\circ}$) was continuously observed from 8:03 UT to 8:20 UT (figure \ref{fig:fig1}).

The horizontal spectrograph of DST is able to cover the entire visible 
and near infrared solar spectrum, and a Universal Tunable Filter 
(hereafter UTF for short, \cite{Hagino2014}) captures the 
slit jaw image at 11 wavelengths ranging from H$\alpha-2.2$ \rm{\AA} 
to H$\alpha+2.2$ \rm{\AA}. 
During the observation, the slit of the spectrograph 
with a width of 100 $\rm{\mu m}$
corresponding to 0.64 arcsec on the sky was fixed to a single position, 
where the eruption and a low altitude stationary prominence 
were covered in its field of view of about 120 arcsec.
They can be seen at the upper and lower part of the spectrum image in figure \ref{fig:fig1}.

In this study, we simultaneously observed He$\,\emissiontype{I}$ 7065 \rm{\AA}, O$\,\emissiontype{I}$ 7772 \rm{\AA} triplet 
and Ca$\,\emissiontype{II}$ 8498 \rm{\AA} using three CCD cameras (Prosilica GE1650, 12-bit) 
with an exposure time of 300 $\rm{ms}$. 
As one camera starts its exposure, trigger signal is sent to the other two cameras, 
with a time lag less than 8 $\rm{\mu s}$. 
This value is much smaller than the typical time scale of seeing at DST, 
which is approximately equal to 1 $\rm{ms}$ with the typical Fried's parameter of 40 $ \rm{mm}$ (\cite{Fried1966}; \cite{Kawate2011}) 
and wind velocity at the turbulent altitude of 40 $ \rm{m\;s^{-1}}$. 
Therefore the exposures of all three cameras can be regarded as exactly simultaneous. 
Time cadence is 0.5 sec for the spectroscopic observation,
while it is 8 sec for the imaging observation by the UTF.

\subsection{Data reduction}

The raw spectral data were subtracted by the averaged dark image taken after the observational sequence. 
Next, the data were corrected for flat field that was made by eliminating absorption lines from the average of about 1000 spectra taken
by scanning quiet regions around the solar disk center. 
Then sky spectrum, which was made by averaging spectral profiles 
over 50 pixels outside the prominence, 
is subtracted after multiplying a factor for adjusting the continuum level
to extract the emission profiles of the eruption.
In the He$\,\emissiontype{I}$ 7065 \rm{\AA} spectral window, 
there are several absorption lines originating from H$_{2}$O molecules 
in the earth's atmosphere, 
and they overlap the extended emission profiles of the He$\,\emissiontype{I}$ line.
Since the depth of these absorption features in the emission profiles
taken through the earth's atmosphere are not identical 
to those in the sky spectrum that is created 
by scattering the continuum in solar spectrum in earth's atmosphere,
they are not eliminated by the sky subtraction.
Therefore, we removed them before sky subtraction 
by dividing narrow regions of spectra including the telluric lines 
in eruption and the sky spectra 
by the corresponding absorption profiles of the telluric lines in the sky spectrum.


The spatial sampling along the slit, spectral sampling 
in units of wavelength and Doppler velocity 
for He$\,\emissiontype{I}$ 7065 \rm{\AA}, O$\,\emissiontype{I}$ 7772 \rm{\AA} triplet and Ca$\,\emissiontype{II}$ 8498 \rm{\AA} 
are listed in Table \ref{tab:table1}. 
Wavelength calibration was performed by comparing the average spectral profile of background sky with the solar atlas (\cite{Moore1966}). 
After resampling in space for He$\,\emissiontype{I}$ and O$\,\emissiontype{I}$ spectra to match with that of Ca$\,\emissiontype{II}$, 
which has the finest spatial sampling, 
alignment was performed among three spectra by using the images 
of two hair lines placed on the entrance slit of the spectrograph.

\begin{table}
  \caption{The spatial sampling along the slit $\rm{\Delta  l_{slit}}$, spectral sampling $\rm{\Delta\lambda}$ in unit of wavelength and Doppler velocity.}
  \label{tab:table1}
    \begin{tabular}{lccc}
      \hline
        & $\rm{\Delta l_{slit}}$ & \multicolumn{2}{c}{$\rm{\Delta \lambda}$} \\
        & $(\rm{arcsec/pix}$) & (\rm{m\AA/pix}) & ($\rm{km/s/pix}$) \\
      \hline
      He$\,\emissiontype{I}$ 7065 \rm{\AA}  & 0.29 & 35 & 1.48 \\
      O$\,\emissiontype{I}$ 7772 \rm{\AA}   & 0.36 & 43 & 1.65 \\
      Ca$\,\emissiontype{II}$ 8498 \rm{\AA} & 0.28 & 31 & 1.10 \\
      \hline
    \end{tabular}
\end{table}

The noise level, which is dominated by the photon noise, 
was evaluated from the standard deviation of the continuum intensity.
The signal-to-noise ratio is obtained as 93, 89 and 27 in the sky continuum
for He$\,\emissiontype{I}$, O$\,\emissiontype{I}$ and Ca$\,\emissiontype{II}$ spectra, respectively. 
The signal-to-noise level in the Ca$\,\emissiontype{II}$ line was relatively low 
because of the lower sensitivity of the camera in this wavelength.

As was done for the spectral data, we performed standard procedures for  
dark and flat corrections on the imaging data taken by the UTF, 
where the flat field was obtained by averaging about 200 images 
taken in nearby continuum of H$\alpha$ by moving the FOV over the solar disk.

\begin{figure*}
 \begin{center}
  \includegraphics[width=18cm]{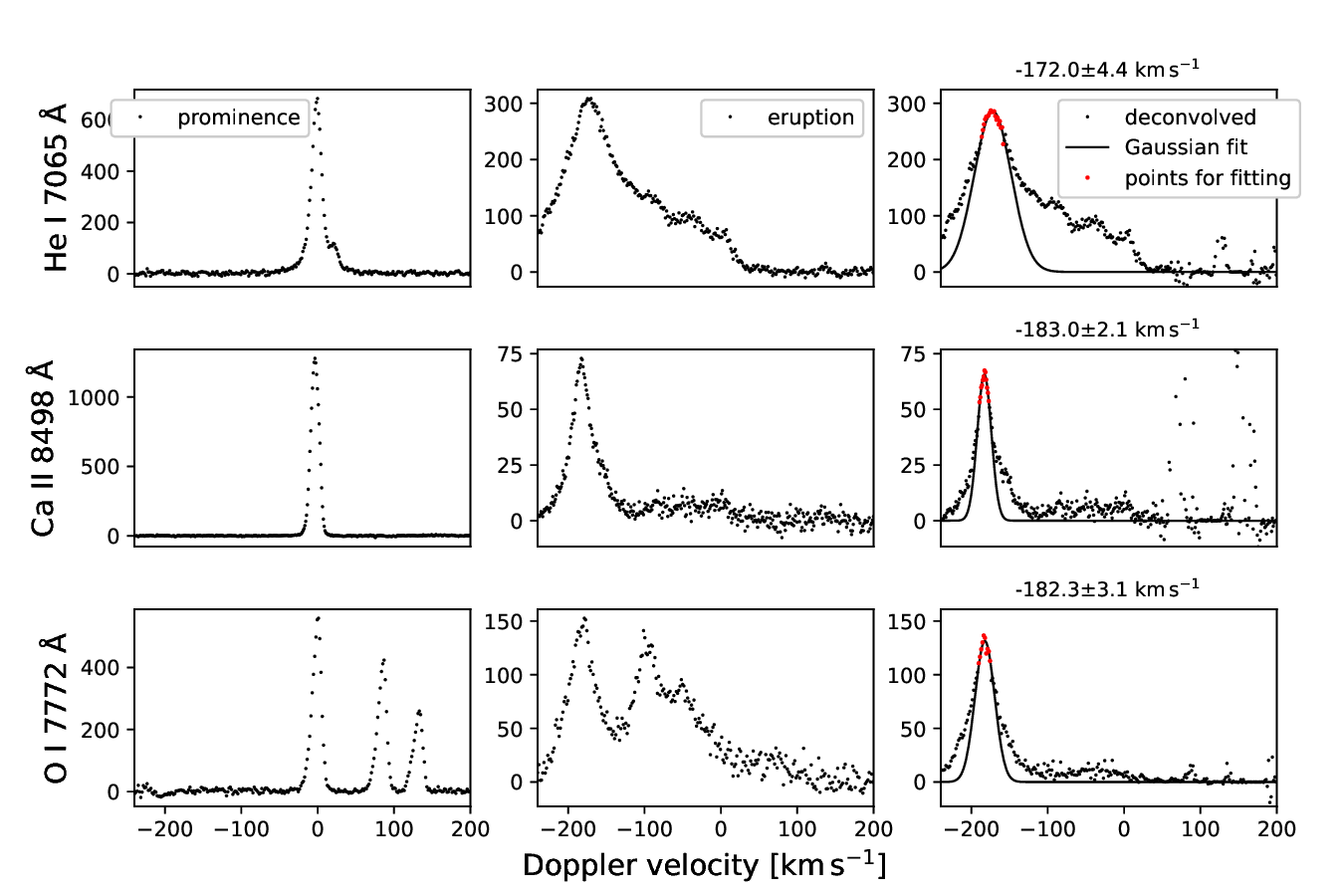} 
 \end{center}
\caption{Left column: the observed profiles of stationary prominence (dot line); middle column: example of averaged observed profiles of the prominence eruption (dotted line); right column: deconvolved profiles (dotted line), 
data points at line core used for Gaussian fitting (red dots) 
and the result of the Gaussian fitting (solid curves).
{Alt text: 9 panels are arranged 3 by 3 in horizontal and vertical.
Each panel shows a spectral profile against the Doppler velocity in km/s.
From the top to the bottom, shown are the spectral profiles 
of He$\,\emissiontype{I}$ 7065 \rm{\AA}, Ca$\,\emissiontype{II}$ 8498 \rm{\AA} and O$\,\emissiontype{I}$ 7772 \rm{\AA}, 
and from the left to the right, those of stationary prominence,
eruption, and deconvolved and fitting profiles.} 
}
\label{fig:fig2}
\end{figure*}

\subsection{Effect of atmospheric differential refraction}
\label{sec:atmospheric-differential-refraction}

The observational setup ensured that data from three spectral windows were taken simultaneously. However, since we are observing with different wavelength around 17:00 in local time with a low solar elevation angle, dispersion caused by the differential refraction in the Earth atmosphere must be taken into account. 

From the latitude of the observatory, solar hour angle and zenith distance, we calculated the direction of dispersion due to differential refraction relative to the slit, by taking the difference between parallactic angle and the inclination of the slit during observation.

To calculate the parallactic angle, we make use of Equation (\ref{eq:eq-parallactic}) (\cite{Makita1996}).

\begin{equation} \label{eq:eq-parallactic}
    \sin{p} = \frac{\cos{\phi}\sin{H_a}}{\sin{Z}},
\end{equation}
where $\phi = 36.25^{\circ}$ is the latitude of the telescope.
$H_a = 78.2^{\circ}$ is the solar hour angle, and $Z = 70.6^{\circ}$ is the zenith distance
at the time of the observation.

From the calculated parallactic angle $p = 56.8^{\circ}$ 
and the inclination angle of the slit,
we found that the angle between the atmospheric dispersion and the slit is about $92.4^{\circ}$, 
which means that the Earth atmospheric layer dispersed light almost perpendicular 
to the direction of the slit. 
Using the empirical formula
and the index of refraction of the atmosphere given by \citet{Owens1967},
we calculated the refraction angle for each wavelength, 
with the local condition at our observation, i.e., 
local pressure $P = 87.1 \,\rm{kPa}$ and temperature $T = 18.0 \rm{^\circ C}$. 
As a result, the atmospheric refraction in 7065 \rm{\AA} and 7772 \rm{\AA} 
relative to 8498 \rm{\AA} are 
obtained as 0.489 and 0.213 arcsec toward the solar limb, respectively.

Therefore, the atmospheric differential refraction is not a negligible issue in our observation, 
which might produces "fake" Doppler velocity difference from their spatial variation. 
We leave the further discussion to section \ref{sec:spatial-variation-of-Doppler-velocity}.

\section{Line Fitting and Doppler velocity}

\subsection{Data selection}

In our study we are interested in establishing the reliable 
Doppler velocity difference between ions and neutrals.
To ensure a sufficient signal-to-noise ratio,
we only processed data where Ca$\,\emissiontype{II}$ spectral profile has 
peak intensity larger than 30 times of the noise level. 
In addition, 
since intensity profiles of the erupting prominence show 
violent features in most cases, 
we excluded profiles with the following criteria 
before performing a fitting procedure,
to avoid erroneous comparison due to the presence of multiple components.

\begin{enumerate}
\renewcommand{\labelenumi}{\arabic{enumi})}
  \item Emission profiles with multiple peaks in the Ca$\,\emissiontype{II}$ line and the O$\,\emissiontype{I}$ lines. 
  In this case, although multiple components in these lines are spectrally resolved 
  and the Doppler velocities could be fitted using multiple Gaussian models in these lines, 
  it is very difficult to find the Doppler velocities belonging to the corresponding 
  components in He$\,\emissiontype{I}$ line. 
  This is because usually these components are not spectrally resolved in He$\,\emissiontype{I}$ line due to its broadness.
  In addition, there might exist more latent components in He$\,\emissiontype{I}$ 7065 \rm{\AA} 
  which are invisible in Ca$\,\emissiontype{II}$ 8498 \rm{\AA} and O$\,\emissiontype{I}$ 7772 \rm{\AA} triplet, 
  since these two lines are intrinsically faint in prominence.
  \item Emission profiles where the inferred Doppler width of the He$\,\emissiontype{I}$ line is 
  larger than 3 times of that of the Ca$\,\emissiontype{II}$ line in velocity.
  In this case, there exist multiple components in He$\,\emissiontype{I}$ line which are invisible in Ca$\,\emissiontype{II}$ line, 
  even if their line core looks like single Gaussian profile.
  
\end{enumerate}

\subsection{Data fitting procedure}
\label{sec:data-fitting-procedure}

After filtering out "bad" profiles with the above criteria, 
the remained profiles were then fitted to obtain the Doppler velocities as follow.

\begin{enumerate}
\renewcommand{\labelenumi}{\arabic{enumi})}
    \item A single snapshot taken at 08:01:29 UT 
    before the eruption happened was selected 
    as reference spectra, from which intensity profiles 
    of the stationary prominence 
    was averaged over 5 pixels along the slit. 
    The reference profiles for three lines are shown in the left column of figure \ref{fig:fig2}.
    After normalization, these averaged profiles were used as kernel functions 
    for deconvolving intensity profiles of the erupting prominence in the following process. 
    Note that these kernels include the PSF (point spread function) 
    of the spectrograph, and intrinsic spectral shapes with multi components of the He$\,\emissiontype{I}$ and O$\,\emissiontype{I}$ lines.
    
    \item For selected intensity profile of the eruption, 
    we averaged it over five pixels along the slit to reduce the noise. 
    Examples of the averaged intensity profiles of the eruption are shown in the middle column in figure \ref{fig:fig2} with dot points, namely "original" profile.
    
    \item The averaged profiles were then deconvolved with their kernel functions using Richardson Lucy deconvolution (\cite{Richardson1972}; \cite{Lucy1974}). 
    Examples of the deconvolved profiles are shown in the right column 
    of figure \ref{fig:fig2} with dot points. 
    The deconvolved profile of O$\,\emissiontype{I}$ 7772 \rm{\AA} now appears as a "singlet", 
    because its kernel function includes all three components.
    
    \item Finally, the deconvolved profiles were fitted with a Gaussian 
    only using the data points in emission core with intensity larger than 80\% 
    of the peak intensity. 
    The fitted Gaussian profiles are illustrated in the right column of figure \ref{fig:fig2} 
    with solid curves. 
    In this step, we estimated the Doppler velocity of the main component of each line 
    in the eruption, by simply taking the value of the center of the Gaussian.
    Note that these kernels, i.e., spectra of the stationary prominence, are our zero reference of the obtained Doppler velocities.
    Here it is assumed that the ion-neutral drift velocity in the stationary prominence is negligible.
    In this way, the speed of Solar rotation was automatically canceled out as well
    so that we are able to avoid the uncertainties produced by the wavelength calibration.
    
\end{enumerate}


Emission profiles are desired to be optically thin to ensure that we are observing the same cloud of plasma through the line-of-sight in all three wavelengths. 
We confirmed that our target clouds are optically thin in He$\,\emissiontype{I}$ 7065 \rm{\AA} from the intensity ratio of its main and sub components. 
On the other hand, although our clouds seem to be optically thin also 
in other two lines from the past prominence observations, 
it is not absolutely evident.
Even if they are not optically thin in  Ca$\,\emissiontype{II}$ 8498 \rm{\AA} and O$\,\emissiontype{I}$ 7772 \rm{\AA}, 
it is not likely that we are observing different clouds in three lines, 
since we carefully selected spectral profiles that do not have multi components with the criteria mentioned in the previous paragraph.
Each target cloud is isolated moving with a high speed as a body.
In such situation, the difference of optical thickness of spectral lines 
will not likely result in the difference of the Doppler velocity.

\begin{figure*}
    \begin{center}
     \includegraphics[width=18cm]{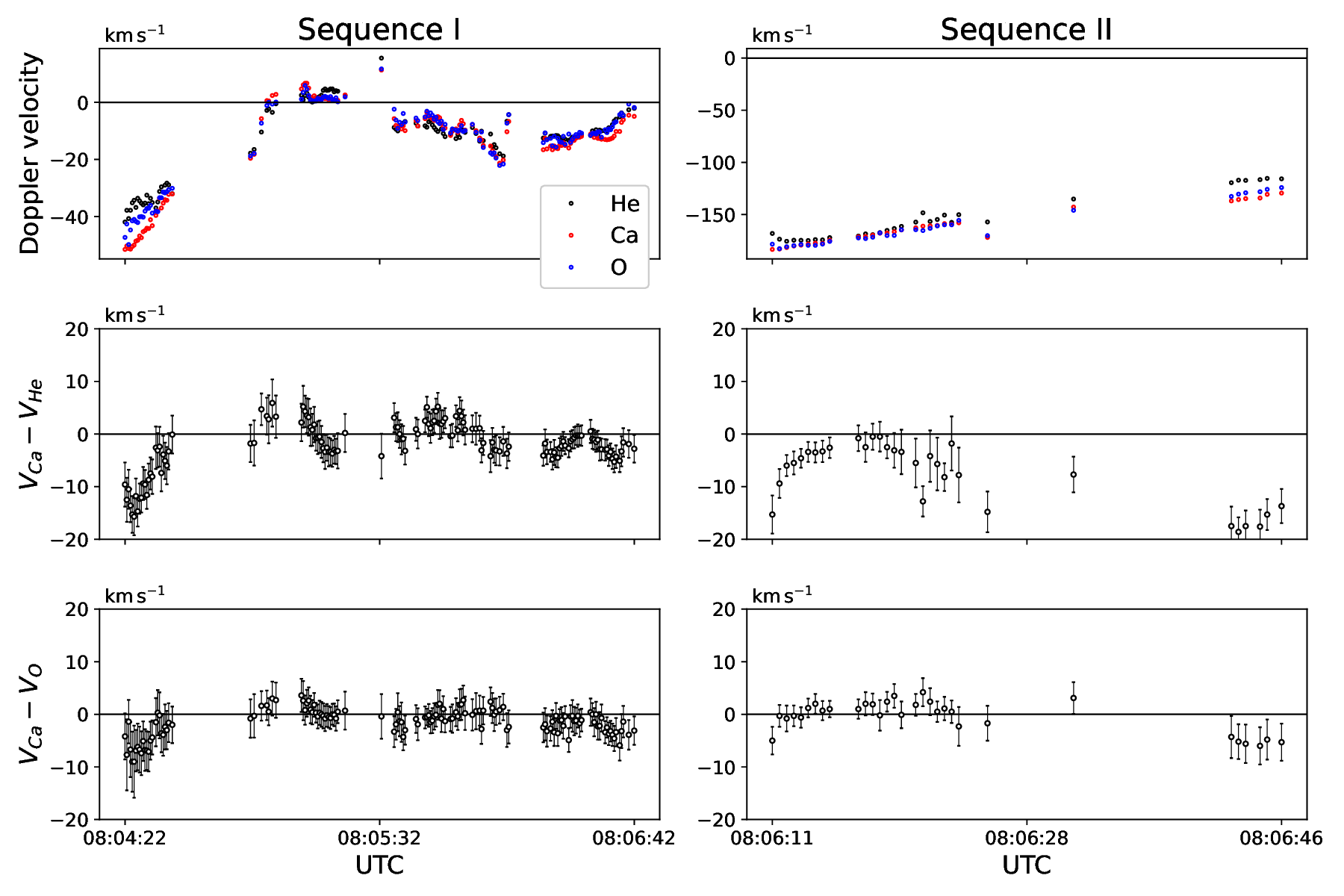} 
    \end{center}
    
    \caption{Time profiles of the Doppler velocity and velocity differences between ion and neutrals for sequence I (left) and sequence II (right). 
    From the top to the bottom, the Doppler velocity, the Doppler velocity difference $V_{\rm{Ca\,\emissiontype{II}}}-V_{\rm{He\,\emissiontype{I}}}$, and the Doppler velocity difference $V_{\rm{Ca\,\emissiontype{II}}}-V_{\rm{O\,\emissiontype{I}}}$. 
    Bars in the velocity difference plots indicate the estimated errors for the Doppler velocity differences.
    {Alt text: Three panels arranged vertically on the left and the right columns,
    which are remarked as Sequence I and Sequence II, respectively.
    In each panel, horizontal axis is the time in UT and vertical axis is velocity in km/s.
    The top panel shows Doppler velocity of He$\,\emissiontype{I}$, 
    Ca$\,\emissiontype{II}$ and O$\,\emissiontype{I}$ lines in different colors. 
    Second and third panels show velocity difference
    of $V_{Ca} - V_{He}$ and $V_{Ca} - V_{O}$, respectively. } 
    }
    \label{fig:fig-seq}
\end{figure*}

\subsection{Error estimation}

Since the intensity profiles in the eruption are much noisier than those 
in the stationary prominence, 
the quality of the fitting process would be more easily affected by the noise. 
We evaluated the error in obtaining the Doppler velocity 
by estimating the shift of the Gaussian by which the fitting residual gets 
comparable with the noise level evaluated from the standard deviation 
in the continuum intensity.
Examples of the fitted error of the Doppler velocities thus obtained  
are shown on the top of each plot in the right column of figure \ref{fig:fig2}.
The error of the difference between two inferred Doppler velocities, $V_1$ and $V_2$ 
was calculated by $\Delta V_{diff} = (\Delta V_1^2 + \Delta V_2^2)^{1/2}$,
where $\Delta$ means the error.

\section{Observational results}

The products of our observation consist of 
(1) time sequence of the Doppler velocities for three lines emanating from He$\,\emissiontype{I}$, O$\,\emissiontype{I}$, and Ca$\,\emissiontype{II}$ and
(2) H$\alpha$ slit-jaw images with their consequent Doppler velocity maps.

\subsection{Doppler velocity}


We obtained time sequences of the Doppler velocity in selected locations on the slit. 
At the beginning of the eruption, 
Ca$\,\emissiontype{II}$ 8498 \rm{\AA} and O$\,\emissiontype{I}$ 7772 \rm{\AA} are faint. 
During the eruption, in most cases, emission lines are strongly shifted, broad and flat, resulting in long tails. 
This is because expansion of the erupting plasma consisted of multiple components
that could not be spatially and spectrally resolved.
At the post-erupting phase, the emission profiles became stable and single Gaussian-like, and Ca$\,\emissiontype{II}$ 8498 \rm{\AA} and O$\,\emissiontype{I}$ 7772 \rm{\AA} became faint again. 

We selected 2 sequences of erupting spectra consisting of 213 frames in total 
that could be reliably analyzed. 
Figure \ref{fig:fig-seq} shows the time variation of the Doppler velocity and its difference between two spectral lines.
The positions of these spectra on the slit are annotated by white arrows in figure \ref{fig:fig1}. 
Note that we are not tracking the motion of the same cloud of plasma but sampling plasma 
successively passing across the slit, although the Doppler velocity in a specific time sequence changed continuously.
For these sequences, we fitted profiles averaged over five pixels 
along the slit around its highest emission intensity at each time step.

Sequence I contains totally 136 fits of spectra taken for a duration of 140 $\rm{s}$ from 08:04:22 to 08:06:42 UT. 
Time variations of the obtained velocity and velocity differences are
shown in the left panel of figure \ref{fig:fig-seq}. 
There are some gaps which can be identified in the velocity and velocity difference plots. 
These gaps correspond to the drops of event selection, 
mainly due to the weak intensity or multi-component like emission core in Ca$\,\emissiontype{II}$ 8498 \rm{\AA}. 
Velocity difference $V_{\rm{Ca\,\emissiontype{II}}}-V_{\rm{He\,\emissiontype{I}}}$ is significant, 
reaching an order of 15 $\rm{km\,s^{-1}}$ at the beginning of the sequence, 
where the Doppler shift is over 40 $\rm{km\,s^{-1}}$. 
After that, Doppler shift decreases and the velocity difference reduces to be 
within the range of the fitting error. 

The right panels of figure \ref{fig:fig-seq} show the plots similar to the left panels 
for another time sequence (Sequence II) with the maximum Doppler velocity larger than 180 $\rm{km\,s^{-1}}$. 
Again, at the beginning of the sequence, 
we clearly see a decay in $V_{\rm{Ca\,\emissiontype{II}}}-V_{\rm{He\,\emissiontype{I}}}$, 
with a maximum velocity difference approaching an order of 15 $\rm{km\,s^{-1}}$. 
In this fast erupting sequence, 
data gaps are mainly due to the insufficiency of intensity in Ca$\,\emissiontype{II}$ 8498 \rm{\AA}.

\begin{figure*}%
 \begin{center}
  \includegraphics[width=18cm]{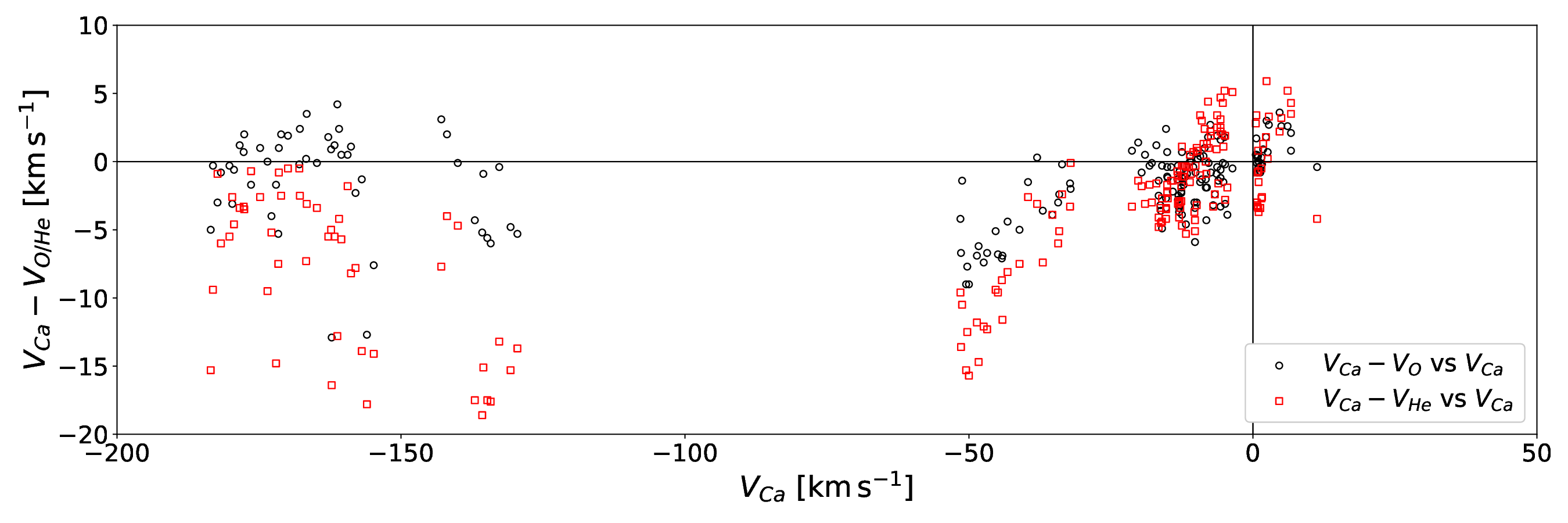} 
 \end{center}
\caption{Scatter plot between $V_{\rm{Ca\,\emissiontype{II}}}$ and $V_{\rm{Ca\,\emissiontype{II}}}-V_{\rm{He\,\emissiontype{I}}}$ (red), $V_{\rm{Ca\,\emissiontype{II}}}-V_{\rm{O\,\emissiontype{I}}}$ (black) of all 213 fitted events.
{Alt text: Velocity difference between ion and neutrals are plotted 
against the Doppler velocity from Ca$\,\emissiontype{II}$ line in km s$^{-1}$.
$V_{Ca} - V_{He}$ and $V_{Ca} - V_{O}$ are shown by red and black symbols, respectively.}
}
\label{fig:fig-stat-a}
\end{figure*}

\begin{figure}
 \begin{center}
  \includegraphics[width=9cm]{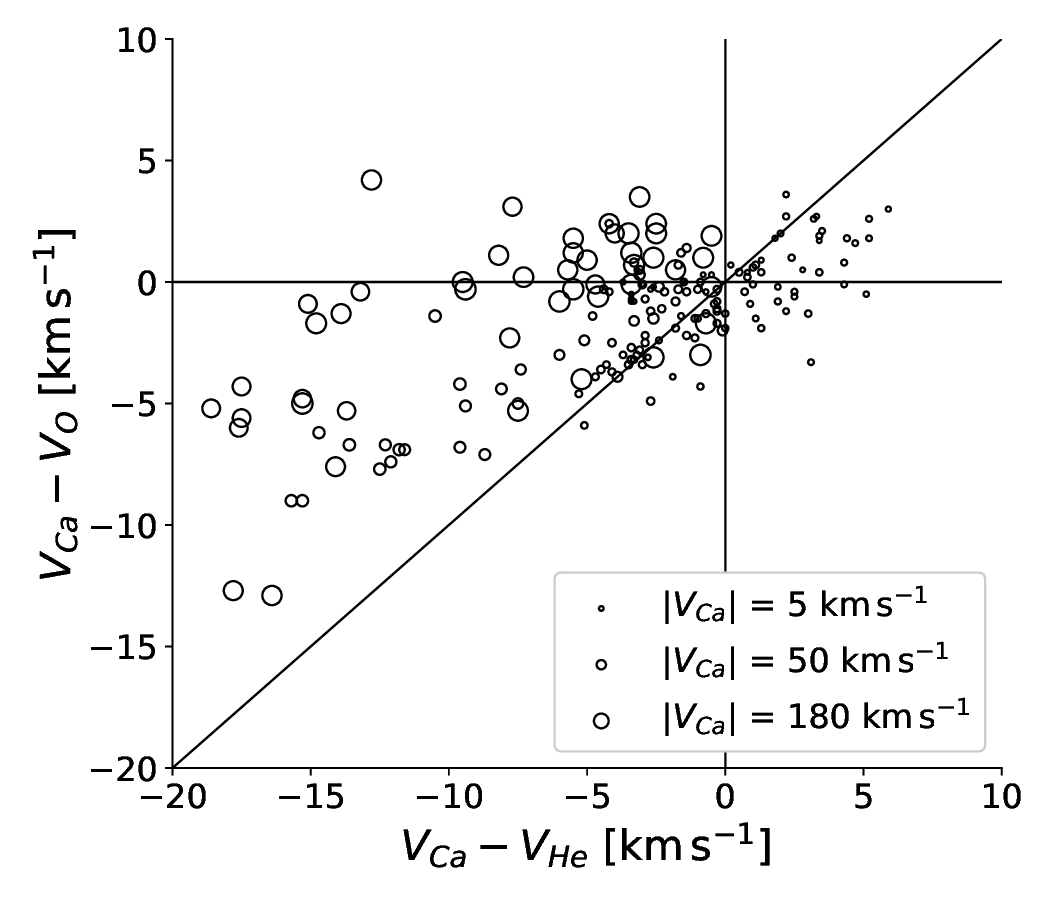} 
 \end{center}
\caption{Scatter plot between $V_{\rm{Ca\,\emissiontype{II}}}-V_{\rm{O\,\emissiontype{I}}}$ and $V_{\rm{Ca\,\emissiontype{II}}}-V_{\rm{He\,\emissiontype{I}}}$ of all 213 fitted events. Marker size is scaled by $|V_{\rm{Ca}}|$.
{Alt text: Horizontal axis is the velocity difference between Ca$\,\emissiontype{II}$ and He$\,\emissiontype{I}$, while vertical axis is the velocity difference 
between Ca$\,\emissiontype{II}$ and O$\,\emissiontype{I}$, each in km s$^{-1}$.}
}
\label{fig:fig-stat}
\end{figure}

It is recognized in the bottom panels of figure \ref{fig:fig-seq}
that the velocity difference $V_{\rm{Ca\,\emissiontype{II}}}-V_{\rm{O\,\emissiontype{I}}}$ is less significant than that of $V_{\rm{Ca\,\emissiontype{II}}}-V_{\rm{He\,\emissiontype{I}}}$.
Figure \ref{fig:fig-stat-a} shows the velocity differences
$V_{\rm{Ca\,\emissiontype{II}}}-V_{\rm{O\,\emissiontype{I}}}$ and $V_{\rm{Ca\,\emissiontype{II}}}-V_{\rm{He\,\emissiontype{I}}}$ against $V_{\rm{Ca\,\emissiontype{II}}}$,
with black and red symbols, respectively.
The groups of symbols in smaller and larger blue shifts correspond to
the sequence I and II, respectively.
It is clear that the He$\,\emissiontype{I}$ 7065 \rm{\AA} (red symbols) exhibits significant
velocity differences with respect to the Ca$\,\emissiontype{II}$ 8498 \rm{\AA},
while the O$\,\emissiontype{I}$ 7772 \rm{\AA} (black symbols) shows much less differences
in both sequences.
Figure \ref{fig:fig-stat} is a scatter plot between two ion-neutral velocity differences in all events, 
where marker size is scaled according to the Doppler velocity of Ca$\,\emissiontype{II}$. 
As we see, with Doppler velocity exceeding 40 $\rm{km\,s^{-1}}$, 
it is more possible to have larger $|V_{\rm{Ca\,\emissiontype{II}}}-V_{\rm{He\,\emissiontype{I}}}|$ reaching to $10$ -- $20$ $\rm{km\,s^{-1}}$, 
while $|V_{\rm{Ca\,\emissiontype{II}}}-V_{\rm{O\,\emissiontype{I}}}|$ rarely goes beyond 10 $\rm{km\,s^{-1}}$.
This result is obviously in contradiction with our earlier prediction 
that heavier neutrals have larger velocity difference from ions.

\begin{figure}
 \begin{center}
  \includegraphics[width=9cm]{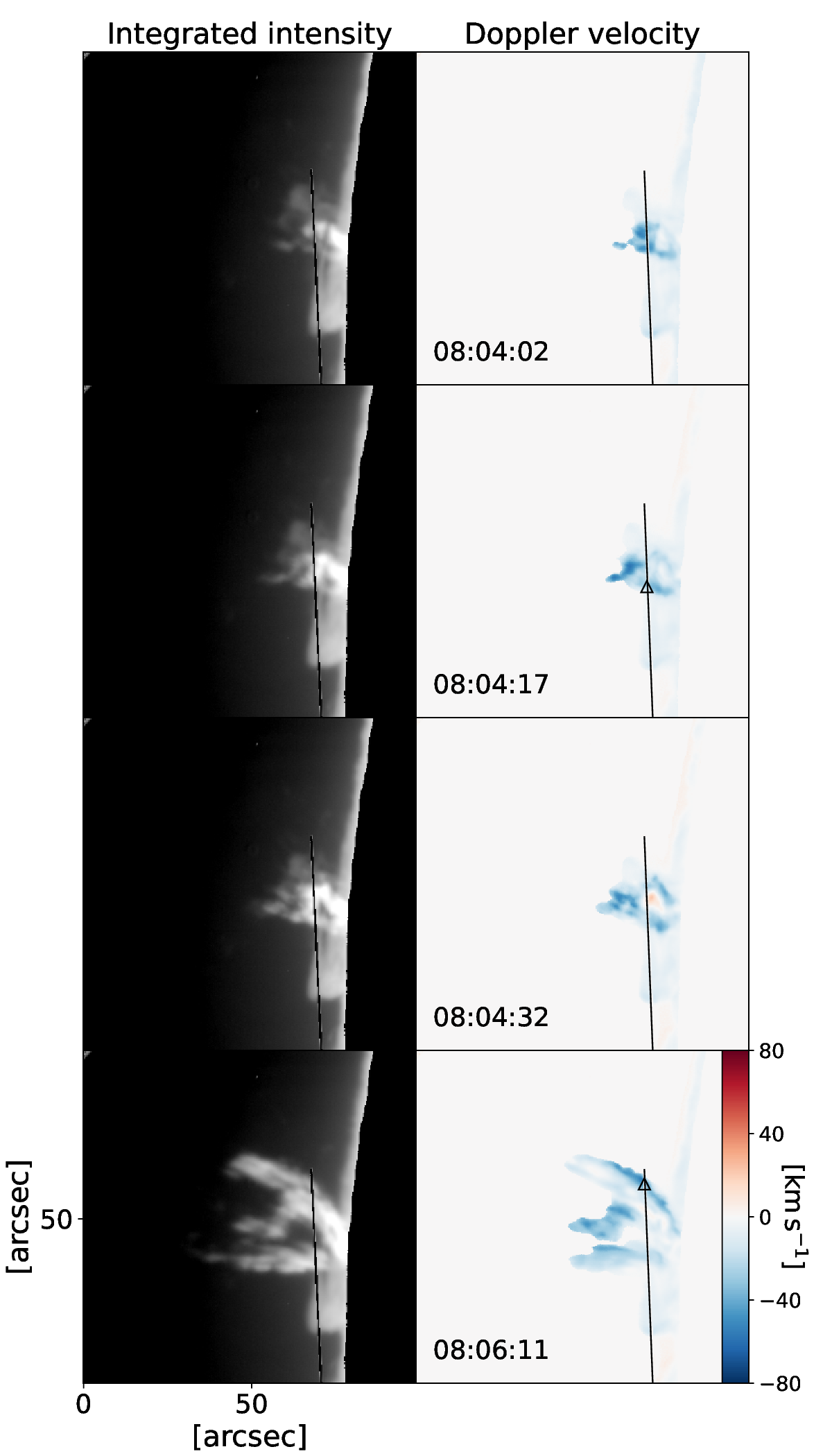} 
 \end{center}
\caption{Left column: $\rm{H\alpha}$ intensity map summing all intensities at the 11 wavelengths in log scale. 
Right column: Doppler velocity map calculated from the center of gravity of the intensity profiles.
Triangles in the second and fourth Doppler velocity maps
point where we extracted the sequences in figure \ref{fig:fig-seq}.
{Alt text: H$\alpha$ images and corresponding Doppler velocity maps 
of the eruption on the left and right at four different times from the top to the bottom.
In each panel, solar limb is seen on the right.}
}
\label{fig:fig-dop-multi}
\end{figure}

\subsection{H$\alpha$ UTF imaging}

Besides the spectral images, we also have slit-jaw images of the eruption 
captured by UTF at 11 wavelengths around $\rm{H\alpha}$, 
in an order of H$\alpha$ $-2.188$, $+2.188$, $-1.094$, $+1.094$, $-0.656$, $+0.656$, $-0.438$, $+0.438$, $-0.219$, $+0.219$, $+0.000$ \rm{\AA}, 
with a time cadence of 8 $\rm{s}$.

Figure \ref{fig:fig-dop-multi} illustrates the growth of the eruption 
in $\rm{H\alpha}$ intensity summed over the all wavelengths in the left column, 
and their corresponding Doppler maps in the right column. 
The straight lines in both columns show the position of our slit for the spectral observation.
Doppler velocity was obtained as the center of gravity of the $\rm{H\alpha}$ profiles.
Here we recognized that the stationary prominence
used as the velocity reference for the spectral data
has a Doppler velocity of about $-3 \rm{km\,s^{-1}}$ in UTF data.
This means that the UTF has a zero-velocity offset with respect to the spectrograph
by about $+3 \rm{km\,s^{-1}}$. 
However this offset has no impact on the following arguments
since we do not use absolute values of the Doppler velocity from UTF.
Note that, since the UTF observation covers the wavelength  
only from -2.2\AA\ to +2.2\AA\ , 
ejecta moving faster than $\pm 100 \rm{km\,s^{-1}}$ in LOS are not
captured in these images.

The top three rows cover the beginning phase of sequence I in figure \ref{fig:fig-seq}, 
where we observed 10--20 $\rm{km s^{-1}}$ velocity difference in $V_{\rm{Ca\,\emissiontype{II}}} - V_{\rm{He\,\emissiontype{I}}}$. 
The position corresponding to sequence I in figure \ref{fig:fig-seq} is marked by a white triangle in the Doppler map in the second row of figure \ref{fig:fig-dop-multi}.
From these three Doppler maps, we found that 
a loop structure broke at the loop top, 
after which plasma around the loop top was accelerated. 
The bottom row shows the appearance after the loop structure totally broke out, 
when the Doppler velocity reached an order of 200 $\rm{km\,s^{-1}}$. 
A white triangle in the Doppler map in the bottom row marks the location 
where we measure sequence II in figure \ref{fig:fig-seq}.

From the time sequences of the wavelength-integrated 
H$\alpha$ images, we evaluated the ascending velocities 
of the eruption on the plane of the sky, 
and they are found to be about 
185$\rm{km\,s^{-1}}$ and 170$\rm{km\,s^{-1}}$ when
the leading edges of ejecta passed the slit in sequences I and II, 
respectively.
By combining with the LOS velocities at the beginning of
sequences I and II, 
it is inferred that these ejecta were lifted off with true velocities of 
$\sim$190$\rm{km\,s^{-1}}$ and $\sim$250$\rm{km\,s^{-1}}$
apart from the sky plane 
by about 15$\degree$ and 45$\degree$ towards us, respectively.

\section{Discussion}

As described in the previous section, we observed significant velocity differences
between Ca$\,\emissiontype{II}$ and He$\,\emissiontype{I}$ in high-velocity components of the erupting prominence.
This might be the first concrete detection of ion-neutral decoupling
in a strongly accelerated plasma.
However, our result unexpectedly shows that the velocity difference
between Ca$\,\emissiontype{II}$ and O$\,\emissiontype{I}$ was less significant than that of Ca$\,\emissiontype{II}$ and He$\,\emissiontype{I}$.
Atmospheric differential refraction also may affect our result.
Therefore we have to make a careful consideration before giving concluding statements.

\subsection{Effect of differential refraction by telluric atmosphere}
\label{sec:spatial-variation-of-Doppler-velocity}

As calculated in section \ref{sec:atmospheric-differential-refraction}, 
the atmospheric differential refraction of 0.213 arcsec between 
Ca$\,\emissiontype{II}$ 8498 \rm{\AA} and O$\,\emissiontype{I}$ 7772 \rm{\AA} is smaller than 
that of 0.489 arcsec between Ca$\,\emissiontype{II}$ 8498 \rm{\AA} and He$\,\emissiontype{I}$ 7065 \rm{\AA}.
Therefore, if there is a large spatial gradient in the line-of-sight velocity 
in the direction of elevation, namely nearly perpendicular to the slit, 
it will result a larger velocity difference between Ca$\,\emissiontype{II}$ and He$\,\emissiontype{I}$ than between Ca$\,\emissiontype{II}$ and O$\,\emissiontype{I}$.

To inspect the spatial variation of Doppler velocity across the slit, 
we make use of a Doppler map taken by UTF. 
We selected a set of H$\rm{\alpha}$ intensity maps taken around 8:04:24 UT, 
at which we measured an order of 10 $\rm{km\,s^{-1}}$ in $|V_{\rm{Ca\,\emissiontype{II}}} - V_{\rm{He\,\emissiontype{I}}}|$,
and a Doppler velocity of about $-40 \, \rm{km\,s^{-1}}$. 
To keep consistent with the Gaussian fitting in our velocity fitting procedure, 
we also fitted the intensity at the 11 wavelength points around H$\rm{\alpha}$ with Gaussian, 
while the Doppler velocity in the stationary prominence was selected as the zero reference, 
as what we did in section \ref{sec:data-fitting-procedure}.

\begin{figure}
 \begin{center}
  \includegraphics[width=9cm]{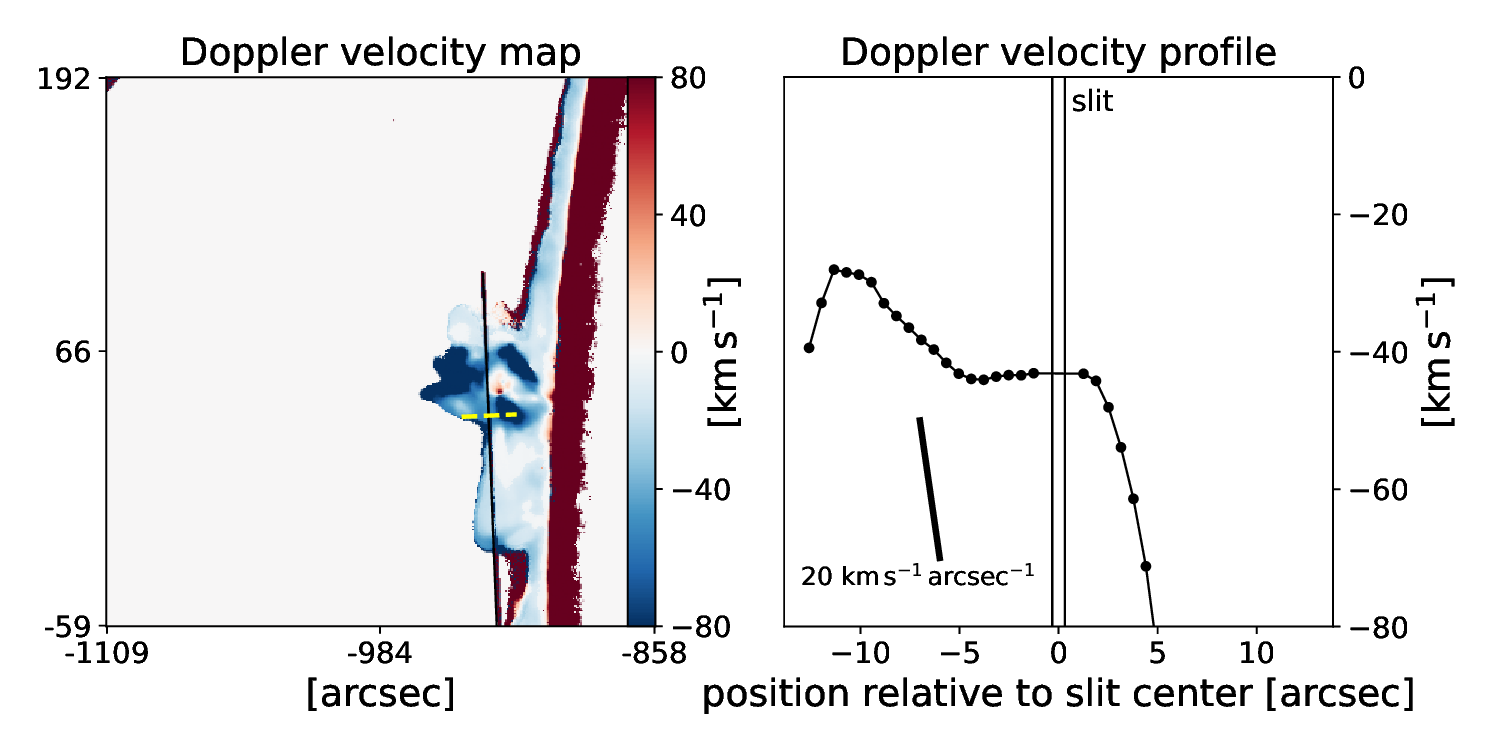} 
 \end{center}
\caption{Left: the Doppler velocity map calculated from UTF H$\rm{\alpha}$ image sequence taken around 8:04:24 UT, where the yellow dashed line is along the direction of atmospheric diffraction. 
Right: the Doppler velocity along the yellow dashed line in the left panel.
The black oblique line in the right panel indicates a velocity gradient, 
which can produce a velocity difference of 10 km s$^{-1}$ between Ca\,\emissiontype{II} 
and He\,\emissiontype{I} due to the atmospheric differential refraction.
{Alt text: Blue-red scale image showing Doppler velocity 
of the eruption is shown on the left.
Yellow dashed line nearly perpendicular to the slit is drawn.
Right panel shows a profile of the velocity along the yellow dashed line in the left panel,
where the position of the slit is indicated.}
}
\label{fig:fig-diff-ref}
\end{figure}

The Doppler velocity map is shown in the left panel of figure \ref{fig:fig-diff-ref}. 
The intersection point between slit and the yellow dashed line is where we extracted 
our Sequence I in figure \ref{fig:fig-seq}.
Along the direction of atmospheric diffraction, which is indicated 
by the yellow dashed line,  
we extracted a profile of the Doppler velocity using cubic interpolation. 
Note that the UTF observation covers the Doppler velocity in the range of [$-90$, $90$] $\rm{km\,s^{-1}}$.
The right panel shows the Doppler velocity along the yellow dashed line 
in the left panel, from which we found that 

\begin{enumerate}
\renewcommand{\labelenumi}{\arabic{enumi})}
    \item the Doppler velocity is about $-40$ $\rm{km\,s^{-1}}$, which is consistent with the result of spectrographic observation in Fig \ref{fig:fig-seq}.
    \item the spatial variation of Doppler velocity is less than $2\,\rm{km\,s^{-1}}/0.5\,\rm{arcsec}$.
\end{enumerate}

Therefore, the velocity difference of $\sim 10 \,\, \rm{km\,s^{-1}}$ between Ca$\,\emissiontype{II}$ and He$\,\emissiontype{I}$ 
obtained from the slit spectra cannot be attributed to the effect of 
differential refraction with a spatial variation of the line-of-sight (LOS) velocity.

\subsection{Formation of O$\,\emissiontype{I}$ 7772 \rm{\AA} triplet}

Since $|V_{\rm{Ca\,\emissiontype{II}}} - V_{\rm{O\,\emissiontype{I}}}| < |V_{\rm{Ca\,\emissiontype{II}}} - V_{\rm{He\,\emissiontype{I}}}|$ cannot be explained 
by the spatial variation of LOS velocity, 
we consider the reason of which by focusing on the formation mechanism 
of the O$\,\emissiontype{I}$ 7772 \rm{\AA} triplet lines.

The infrared triplet lines of O$\,\emissiontype{I}$ (7771.9, 7774.2, 7775.4 \rm{\AA}) are
the most studied spectral lines of neutral oxygen in solar-type stars, 
and it has been known that these lines are formed under non local thermodynamic equilibrium
(non-LTE) conditions in the solar photosphere \citep{Altrock1968}.
The ionization potential of neutral oxygen from the ground state
(13.62 $\rm{eV}$) is close to the ionization potential of neutral hydrogen 
from its ground state (13.60 $\rm{eV}$).
As a result, charge transfer process between oxygen and hydrogen atoms
$\mathrm{H\,\emissiontype{II} + O\,\emissiontype{I} \Longleftrightarrow H\,\emissiontype{I} + O\,\emissiontype{II}}$ 
takes an important role on the state population in the statistical equilibrium.
The rate of charge transfer from a neutral oxygen to proton (ionization of oxygen)
and from neutral hydrogen to an oxygen ion (recombination of oxygen) 
are given by $\nu_i \sim \sigma_{CT} n_p V_{th}$ 
and $\nu_r \sim \sigma_{CT}$$ n_{H\,\emissiontype{I}}$$ V_{th}$, respectively.
Here $n_p$ and $n_{H\,\emissiontype{I}}$ are the number densities of proton and neutral hydrogen,
$V_{th}$ is the mean thermal velocity of proton 
given by $\sqrt{8k_BT/\pi m_p}$, and 
$\sigma_{CT}$ is the charge transfer cross section 
between oxygen and hydrogen atoms,
which is given as $10^{-15}$ cm$^2$ for both directions of charge transfer
in chromospheric condition \citep{Stancil1999}. 
Denoting the number densities of neutral and ionized oxygen 
as $n_{O\,\emissiontype{I}}$ and $n_{O\,\emissiontype{II}}$, respectively, 
and taking into account the balance between 
ionization and recombination processes of the oxygen,
i.e., $\nu_i n_{O\,\emissiontype{I}}$$=\nu_r n_{O\,\emissiontype{II}}$, then
$n_{O\,\emissiontype{II}}$$/n_{O\,\emissiontype{I}}$$=\nu_i/\nu_r \sim n_p/n_{H\,\emissiontype{I}}$.
If we take $n_p \sim n_{e} \sim 10^{10}$ cm$^{-3}$ \citep{Hashimoto2023} and 
$V_{th}\sim 10^6$ cm s$^{-1}$ as typical condition of prominences,  
$\nu_i$ is in the order of $10$ sec$^{-1}$.
This means that the ionization fraction of the oxygen gets comparable 
to that of the hydrogen in a time scale of 0.1 sec.

According to \cite{Pazira2017}, 
the lower level population of the O$\,\emissiontype{I}$ 7772 \AA\ transition 
is mainly regulated through recombination cascades from O$\,\emissiontype{II}$,
and the 7772 \AA\ lines are then generated by scattering 
of the photospheric radiation.
Since the timescale of the ionization-recombination cycle of oxygen
would be shorter than the MHD dynamical timescale,
oxygen atoms stay as ion for a fraction of time comparable to 
the ionization degree of hydrogen during the acceleration.
For example, if the ionization degree of hydrogen is 80\%,
the oxygen atoms stay as ion in 80\% of time.
Therefore, oxygen atoms occasionally feel the Lorentz force and
O$\,\emissiontype{I}$ 7772 \rm{\AA} triplet emission lines behave 
more like ionic lines resulting in 
$|v_{\rm{Ca\,\emissiontype{II}}} - v_{\rm{O\,\emissiontype{I}}}| 
< |v_{\rm{Ca\,\emissiontype{II}}} - v_{\rm{He\,\emissiontype{I}}}|$.
In this way, the degree of decoupling between the neutral oxygen and ion is
related to the ionization degree of hydrogen.

It is remarkable that \citet{Wiehr2025} termed such effect as 'ionization memory'.
They observed almost equal non-thermal broadening 
of lines from ionized Fe and neutral Na and Mg in a prominence,
and argued that the excess broadening of the neutral atoms is partly due to a result of 
their previous existence as ions.

\subsection{Ionization degree}

\begin{figure}
 \begin{center}
 \includegraphics[height=5cm]{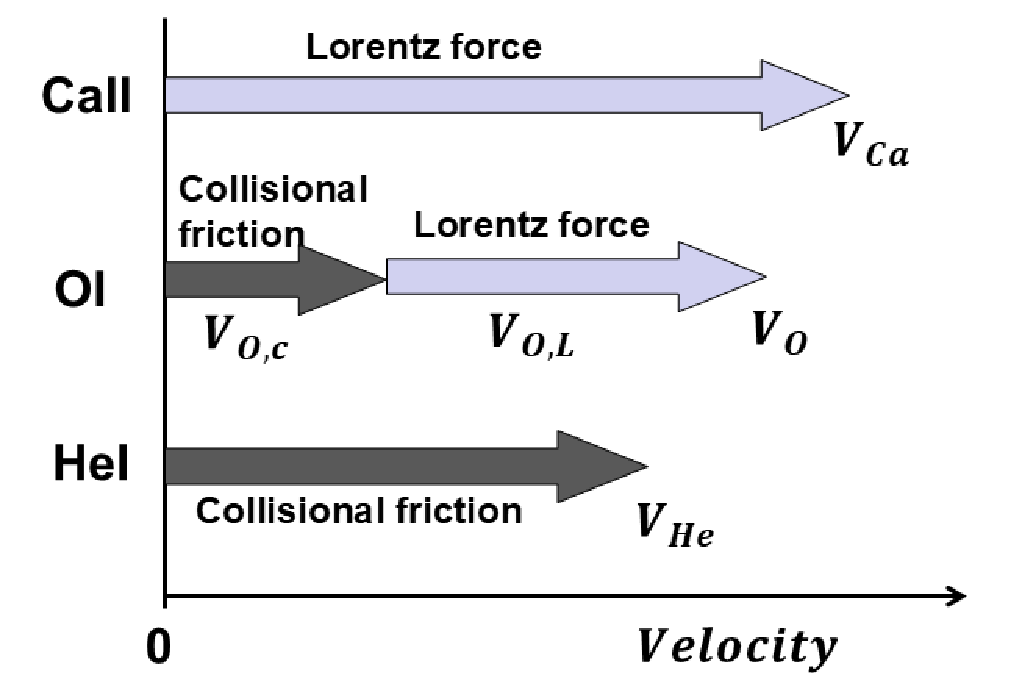} 
 \end{center}
\caption{
Concept of generating the velocity difference of the three atomic lines.
Each arrow shows velocity gain from the Lorentz force or collisional friction. 
{Alt text: Three set of horizontal arrows directed to the right are arranged vertically
on the horizontal axis of velocity.
From the top to the bottom, length of the arrows show the velocity gain 
for CaII, OI and HeI, respectively.
Contributions from the Lorentz force and the collisional friction are displaied 
by different gray scales of the arrows.}
}
\label{fig:fig_V_ionization}
\end{figure}

Taking advantage of the aforementioned relation, we can consider a new method for evaluating 
the ionization degree of hydrogen from the velocity differences observed 
in Ca$\,\emissiontype{II}$, O$\,\emissiontype{I}$ and He$\,\emissiontype{I}$.
Figure \ref{fig:fig_V_ionization} schematically shows how the three atomic species gain
the velocity from the Lorentz force and collisional friction.
We denote the observed velocities of Ca$\,\emissiontype{II}$, O$\,\emissiontype{I}$ and He$\,\emissiontype{I}$ 
as $V_{Ca}$, $V_{O}$ and $V_{He}$, respectively.
Ca$\,\emissiontype{II}$ atoms gain the velocity from the Lorentz force and 
He$\,\emissiontype{I}$ atoms gain the velocity from the collisional friction throughout the acceleration process.
On the other hand, oxygen atoms gain the velocity from both collisional friction and Lorentz force through ionization and recombination cycles.

Now let's recall the equation of motion of neutral atoms 
introduced in Section 1, i.e., $\nu m_p \Delta V = m_n a$.
We can rewrite this as $ F_c = m_n a_n = m_n \delta V_n / \delta t$,
where, $F_c$ is the collisional friction, $a_n$ is the acceleration 
and $\delta V_n$ is the velocity gain of the neutral atoms 
from the collisional friction in a time period of $\delta t$.
This relation reduces to $\delta V_n = F_c m_n^{-1} \delta t$, 
namely, the velocity gain of neutral atoms in a given time is inversely proportional 
to the atomic mass.
It means that, if oxygen atoms feel the collisional friction throughout the acceleration process,  
the velocity gain of oxygen from the collisional friction, $V_{O,C}$, 
is smaller than that of He$\,\emissiontype{I}$ by the ratio of their atomic mass, 
i.e., $m_{He}/m_{O}=1/4$.
Therefore, the velocity gain of oxygen from the Lorentz force, $V_{O,L}$, is equal to 
$V_{O}-V_{O,C} = V_{O} - V_{He}/4$.
Since the velocity gain by the Lorentz force is proportional 
to the time duration of its action, 
the ratio of $V_{O,L}/V_{Ca}$ gives the fraction of time interval
by which oxygen atoms stay as ion, namely, the ionization degree of hydrogen.
For example, if we evaluate $V_{Ca}$, $V_{O}$ and $V_{He}$ 
as 50 km s$^{-1}$, 40 km s$^{-1}$ and 35 km s$^{-1}$, respectively, 
from our observation (Sequence I in figure \ref{fig:fig-seq}),
we obtain the ionization degree of hydrogen as 60\%.

\subsection{He$\,\emissiontype{I}$ - ion drift velocity}

Contrary to the oxygen, helium atoms likely stay as neutral
during the eruption due to its high ionization potential (24.6 eV).
Let's recall the simple model of a partially ionized plasma introduced in Sec.1,
Now it is composed of neutral helium and proton, 
and the momentum balance equation is written as
$\nu m_p \Delta V = m_{He} a$,
with $\Delta V$ the velocity difference between He$\,\emissiontype{I}$ and Ca$\,\emissiontype{II}$,
$a$ acceleration and $\nu$ the collision frequency between 
neutral helium and proton.
In this case, $\nu = \sigma_{MT}$$ n_p v_{th}$
with $\sigma_{MT}$ the momentum transfer cross section
between neutral helium and proton, and it is given by \citet{Vranjes2013}
as $\sim 10^{-15}$ cm$^2$ for the proton energy of 1 eV.
By assuming $n_p=10^{10}$cm$^{-3}$ and $v_{th}=10^6$cm s$^{-1}$ as before,
we can get rough evaluation of the acceleration $a$ from the observed $\Delta V$.
By adopting $\Delta V=15$ km s$^{-1}$, we get $a\sim 38$ km s$^{-2}$ or
about 150 times of $g_\odot$ (gravitational acceleration on the solar surface).
Note that this result is consistent in scale with \citet{Gilbert2002},
in which they derived the drift velocity of He$\,\emissiontype{I}$ against ions
as $\sim 0.1$ km s$^{-1}$ under 1 $g_\odot$ acceleration.

It is worth to compare our result with past studies on the solar eruption.
For example, \citet{Cabezas2024} obtained an acceleration of 2.5 km s$^{-2}$ 
for a high speed eruption using H$\rm{\alpha}$ imaging spectroscopic data.
Our result is 15 times larger than their result.
On the other hand, \citet{Ying2018} obtained $\sim$50 km s$^{-2}$ 
in an ejection associated with a M-class flare
but in high temperature plasma component observed by AIA 93 \AA.
The large acceleration in cool plasma obtained in this study 
may imply a blast impact such as the action of fast MHD shock,
but the detailed mechanism remains a subject for further investigation.


\section{Conclusion}

With a spectroscopic observation of a violent eruption on the solar limb
in He$\,\emissiontype{I}$ 7065 \rm{\AA}, O$\,\emissiontype{I}$ 7772 \rm{\AA} triplet 
and Ca$\,\emissiontype{II}$ 8498 \rm{\AA} lines, 
we found a significant velocity difference between Ca$\,\emissiontype{II}$ and He$\,\emissiontype{I}$ lines 
reaching about 15 km s$^{-1}$.
This could be the first concrete evidence of ion-neutral decoupling 
observed in highly accelerated solar plasma.
On the other hand, O$\,\emissiontype{I}$ lines, which are initially expected to show
a larger velocity difference from ion due to its larger atomic mass,
had significantly smaller velocity differences from Ca$\,\emissiontype{II}$ ion.
The reason of which is interpreted as oxygen atoms were in ionized state
in significant fraction of time during the eruption due to 
the efficient charge exchange with hydrogen and felt the Lorentz force
before emitting the O$\,\emissiontype{I}$ 7772 lines.
From the drift velocity between He$\,\emissiontype{I}$ and Ca$\,\emissiontype{II}$ and simple model 
of collisional friction between neutral helium and proton, 
we estimated the acceleration of the eruption 
to be about 150 times of the solar gravity.
The mechanism to realize this very large acceleration is a subject for future study.




\begin{ack}
The authors are grateful to Mr. Satoru Ueno and staff member of
Hida observatory for supporting our observation at DST.
We also thanks to Dr. Tomoko Kawate for valuable comments on the
plasma physics.
\end{ack}


\begin{thebibliography}{}
\bibitem[Altrock (1968)]{Altrock1968}
  Altrock, E. C. \ 1968, Sol. Phys., 5, 260


\bibitem[Anan et al. (2017)]{Anan2017}
  Anan, T., Ichimoto, K., Hillier, A. \ 2017, A\&A, 601, 103
  
\bibitem[Cabezas et al. (2024)]{Cabezas2024}
  Cabezas, D. P., Ichimoto, K., Asai, A., et al. \ 2024, A\&A, 690, A172 

\bibitem[Fried (1966)]{Fried1966}
  Fried, D. L. \ 1966, J. Opt. Soc. Amer., 56, 1372

\bibitem[Gilbert et al. (2002)]{Gilbert2002}
  Gilbert,H. R., Kilper, G., and Alexander, G., \ 2002, ApJ, 577, 464

\bibitem[Gilbert et al. (2007)]{Gilbert2007}
  Gilbert,H. R., Hansteen, V. H., and Holzer, T. E., \ 2007, ApJ, 671, 978

\bibitem[Hagino et al. (2014)]{Hagino2014}
  Hagino, M., Ichimoto, K., Kimura, G., Nakatani, Y., Kawate, T., Shinoda, K., Suematsu, Y., Hara, H., Shimizu, T. \ 2014, SPIE, 9151, 915115
  
\bibitem[Hashimoto et al. (2023)]{Hashimoto2023}
  Hashimoto, Y., Ichimoto, K., Huang, Y. \ 2023, PASJ, 75, 913

\bibitem[Hillier et al. (2016)]{Hillier2016}
  Hillier, A., Takasao, S., Nakamura, N. \ 2016, A\&A, 591, A112
  
\bibitem[Kawate et al. (2011)]{Kawate2011}
  Kawate, T., Hanaoka, Y., Ichimoto, K., Miura, N. \ 2011, MNRAS, 416, 2154

\bibitem[Khomenko \& Collados (2012)]{Khomenko2012}
  Khomenko, E., Collados, M. \ 2012, ApJ, 747, 87

\bibitem[Khomenko et al. (2014)]{Khomenko2014}
  Khomenko, E., Diaz, A., de Vicente, A., Collados, M., Luna, M. \ 2014, A\&A, 565, A45
  
\bibitem[Khomenko et al. (2016)]{Khomenko2016}
  Khomenko, E., Collados, M., Diaz, A. \ 2016, ApJ, 823, 132

\bibitem[Leake et al. (2012)]{Leake2012}
  Leake, J. E., Vyacheslav, S. L., Linton, M. G., Meier \ 2012, ApJ, 760, 109

\bibitem[Lucy (1974)]{Lucy1974}
  Lucy, L., B. \ 1974, Astronomical Journal, 79, 745
  
\bibitem[Moore et al. (1966)]{Moore1966}
  Moore, C. E., Minnaert, M. G., Houtgast, J. \ 1966, The Solar Spectrum 2935 \rm{\AA} to 8770 \rm{\AA} (Washington: US Government Printing Office (U5GPO))
  
\bibitem[Makita et al. (1996)]{Makita1996}
  Makita, M., Funakoshi, Y., Tomura, I., Kawakami, S., Hanaoka, Y., Kawai, G., \ 1996, Technical Reports from Kwasan and Hida Observatories Faculty of science, Kyoto University, 7: 1-27

\bibitem[Nakai \& Hattori (1985)]{NakaiHattori1985}
  Nakai, Y., Hattori, A. \ 1985, Memoirs of the Faculty of Science, Kyoto University, 36, 385
  

\bibitem[Owens (1967)]{Owens1967}
  Owens \ 1967, Applied Optics, 6, 51

\bibitem[Pazira et al. (2017)]{Pazira2017}
  Pazira, H., Kiselman, D. and Leenaarts, J.  \ 2017, A\&A 604, A49 
  
\bibitem[Richardson (1972)]{Richardson1972}
  Richardson, W., H. \ 1972, JOSA, 62, 55-59
  
\bibitem[Stancil et al. (1999)]{Stancil1999}
  Stancil, P., C., Schultz, D., R., Kimura, M., Gu, J., P., Hirsch, G., Buenker, R., J. \ 1999, A\&A, 140, 255

\bibitem[Snow \& Hillier (2020)]{Snow2020}
  Snow, B., Hillier, A. \ 2020, A\&A, 637, A97

\bibitem[Tomida et al. (2015)]{Tomida2015}
  Tomida, K., Okuzumi, S., Machida, M. N. \ 2015, ApJ, 801, 117

\bibitem[Vranjes and Krstic (2013)]{Vranjes2013}
  Vranjes, J., Krstic, P.S. \ 2013, A\&A, 554, A22

\bibitem[Wiehr et al. (2021)]{Wiehr2021}
  Wiehr, E., Stellmacher, G., Balthasar, H., and Bianda, M., \ 2021, ApJ, 920, 47

\bibitem[Wiehr et al. (2025)]{Wiehr2025}
  Wiehr, E., Balthazar, H., Stellmacher, G., and Bianda, M., \ 2025, A\&A, 696, A209

\bibitem[Ying et al. (2018)]{Ying2018}
Ying, B., Feng, Li., Lu,L. et al. 2018 ApJ, 856, 24        

\bibitem[Zapior et al. (2022)]{Zapior2022}
  Zapior, M., Heinzel, P., and Khomenko, E., \ 2022, ApJ, 934, 16


\end{thebibliography}

\end{document}